\documentclass[a4paper]{jpconf}
\usepackage{graphicx}
\begin{document}
\title{High sensitivity tests of the Pauli Exclusion Principle with VIP2}
\author{J.~Marton$^1$, S.~Bartalucci$^2$, S.~Bertolucci$^4$, C.~Berucci$^{1,2}$,
  M.~Bragadireanu$^{2,3}$, M.~Cargnelli$^1$, C. Curceanu$^{2,3}$, A.Clozza$^2$, S.~Di~Matteo$^5$, J.-P.~Egger$^6$,
  C.~Guaraldo$^2$,  M.~Iliescu$^2$, T.~Ishiwatari$^1$, M.~Laubenstein$^7$,
E.~Milotti$^8$, A.~Pichler$^1$, D.~ Pietreanu$^2$, K.~Piscicchia$^{2,9}$, T.  Ponta$^{2,3}$, A.~Scordo$^2$, H.~Shi$^2$,
D.L.~Sirghi$^{2,3}$, F. Sirghi$^{2,3}$, L.~Sperandio$^2$, O.~Vazquez Doce$^2$, E.~Widmann$^1$ and J.~Zmeskal$^1$}

\address{$^1$ Stefan Meyer Institute for Subatomic Physics, Austrian Academy of Sciences, Boltzmanngasse 3, A-1090 Vienna, Austria}

\address{$^2$ INFN, Laboratori Nazionali di Frascati, CP 13,
 Via E. Fermi 40, I-00044, Frascati (Roma), Italy}

\address{$^3$ "Horia Hulubei" National Institute of Physics and
 Nuclear Engineering, \\ Str. Atomistilor no. 407, P.O. Box MG-6,
Bucharest - Magurele, Romania}

\address{$^4$ CERN, CH-1211, Geneva 23, Switzerland}

\address{$^5$ Institut de Physique UMR CNRS-UR1 6251, Universit\'e de Rennes1, F-35042 Rennes, France}

\address{$^6$ Institut de Physique, Universit\'e de Neuch\^atel, 1 rue A.-L. Breguet, CH-2000 Neuch\^atel, Switzerland}

\address{$^7$ Laboratori Nazionali del Gran Sasso, S.S. 17/bis, I-67010 Assergi (AQ), Italy}

\address{$^8$ Dipartimento di Fisica, Universit\`{a} di Trieste and INFN-- Sezione di Trieste, Via Valerio, 2, I-34127
Trieste, Italy}

\address{$^9$ CENTRO FERMI, Compendio del Viminale, Piazza del Viminale 1, I-00184 Roma, Italy}

\ead{johann.marton@oeaw.ac.at}

\begin{abstract}
The Pauli Exclusion Principle is one of the most fundamental rules of nature and represents a pillar of modern physics.
According to many observations the Pauli Exclusion Principle must be extremely well fulfilled.
Nevertheless, numerous experimental investigations were performed to search for a small violation of this principle.
The VIP experiment at the Gran Sasso underground laboratory searched for Pauli-forbidden X-ray transitions in copper atoms  using the Ramberg-Snow method and
obtained the best limit so far. The follow-up experiment VIP2 is designed to reach even higher sensitivity. It
aims to improve the limit by VIP by orders of magnitude.
The experimental method, comparison of different PEP tests based on different assumptions and the developments for VIP2 are presented.

\end{abstract}

\section{Introduction}

The Pauli Exclusion Principle (PEP) predicates that two fermions (having half-integer spin) cannot occupy the same quantum state - unlike bosons having integer spin.
According to the present knowledge it is a consequence of the spin-statistic theorem, however the physics background remains an open topic.
Wolfgang Pauli introduced PEP and published it in a famous paper \cite{pauli25} ninety years ago. In his nobel
lecture \cite{pauli46} he confessed that a intuitive explanation for PEP he was unable to give:\\

{\it Already in my original paper I stressed the circumstance that I was unable to
give a logical reason for the exclusion principle or to deduce it from more
general assumptions. I had always the feeling and I still have it today, that
this is a deficiency.}\\
\\
Nevertheless he  tried to prove it on the basis of rather complicated arguments \cite{pauli40}. Our understanding is that there are only two spin-separated classes of particles and compound systems namely fermions and bosons in nature, which are
characterized by the half-integer and integer spin respectively. According to many observations PEP is extremely well-fulfilled
for all fermions like nucleons and electrons. The PEP represents a cornerstone of quantum mechanics and has many consequences, e.g. periodic system of the elements, stability of matter, existence of compact stellar objects like neutron stars.
Because of the fundamental importance of PEP speculations about tiny violations were raised. Many experimental attempts based on various assumptions concerning PEP violation were carried out to test the validity with higher and higher sensitivity. Very stringent limits were found for Pauli-forbidden reactions in stable systems by the experiments DAMA  \cite{bernabei97, bernabei09} and Borexino  \cite{borexino05, borexino10} at the Gran Sasso Laboratory. However, according to the so-called Greenberg-Messiah superselection rule \cite{messiah64} reactions between different symmetry classes are forbidden in a stable system with established symmetry. Therefore, this results in an absence of Pauli-forbidden processes in stable systems.
In a pioneering experiment Ramberg and Snow (RS) \cite{ramberg90} used an electric current to provide "new" fermions (i.e. distant originating fermions which have no connections to the atoms under test) to test PEP. The Pauli-forbidden X-ray transitions in copper (K$\alpha$) exhibit an energy shift of about 300 eV relative to the normal 2p-1s transitions (7.729 keV instead of 8.040 keV) \cite{sperandio06}. Differently to searches in stable systems this method avoids the Greenberg-Messiah superselection rule. The upper limit obtained in the Ramberg-Snow experiment was

\begin{equation}\label{eq1}
  \beta^2/2 < 1.7 \cdot 10^{-26}
\end{equation}

The quantity $\beta^{2}$/2 stands for the probability of PEP violation which can be traced back to an theoretical attempt by Ignatiev and Kuzmin \cite{ignatiev87} to formulate PEP violation which was afterwards discarded because of negative probabilities. Nevertheless $\beta^{2}$/2 is widely used in the literature as measure for bounds of PEP violation and will be used also in this paper.
New attemps to accommodate spin-statistics violation were performed in the framework of superstring theory \cite{jackson08} and spacetime non-commutativity \cite{balachandran10}.

\section{VIP experiment at LNGS}

The VIP experiment (Violation of the Pauli Principle) \cite{vip} was designed to refine and improve the search for PEP forbidden X-ray transitions using the Ramberg-Snow method. The main improvements were the use of X-ray detectors with much better energy resolution and a low-background location for the experiment. As X-ray detectors charge coupled devices (CCDs) \cite{egger} were employed which were successfully used in experiments on kaonic atoms at DAFNE first \cite{DEAR,ishiwatari}. Compared to the proportional tube detectors used in the RS experiment the CCDs exhibit superior energy resolution (about 5-times better than the detectors in the RS experiment). The VIP experiment was installed in the underground laboratory Gran Sasso which provides a suppression of the cosmic ray flux by 10$^{6}$ and is therefore ideally suited for low-counting experiments. The VIP experiment used an electric current of 40 A through an ultrapure copper cylinder to introduce "new" electrons for testing PEP. Data were collected with current and without current to search for a possible difference caused by PEP violation. The difference in the expected counts for the 2 data sets comes from the fact, that in the case without current, non-Paulian transitions could have already occured in a long time span since atomic formation. As the current introduces "new" electrons to the system, the non-Paulian transitions could occur during the measurement. In the first data taking period we succeeded to improve the $\beta^{2}$/2 limit of Ramberg-Snow by nearly 2 orders of magnitude (see tab. \ref{VIPresults}).

\begin{table}
\caption{Limits  of the Pauli violation probability for electrons from different experiments using the RS method.}
\label{VIPresults}
\begin{center}
\begin{tabular}{llll}
\br
Experiment&Target&Upper limit of $\beta^2$/2 & ref.\\
\mr
Ramberg-Snow& Copper& 1.7x10$^{-26}$& \cite{ramberg90}\\
S.R. Elliott et al.&Lead&1.5x10$^{-27}$ & \cite{elliott}\\
VIP(2006)&Copper&4.5x10$^{-28}$ & \cite{vipart}\\
VIP(2012)&Copper&4.7x10$^{-29}$ & \cite{laura, curceanu11}\\
VIP2(goal)&Copper&10$^{-31}$ & \\
\br
\end{tabular}
\end{center}
\end{table}

In the extended data taking periods at LNGS the limit was again improved to the region of 10$^{-29}$. The final analysis is finished now and a publication of the final result is in preparation.

\section{The new experiment VIP2}

\subsection{Goal of VIP2}

After the successful completion of VIP we want to further increase the sensitivity for PEP violation in the electron sector. In order to gain several orders of magnitude a new experimental setup with different improvements is needed \cite{Hexi14,Curceanu14,Pisicchia15,Pisicchia15a}.

\begin{itemize}
  \item Enhancement of the acceptance of the X-ray detector
  \item Usage of higher current
  \item Background suppression by smaller target and active shielding
\end{itemize}

\begin{figure}[h]
\centering
\includegraphics[width=15pc]{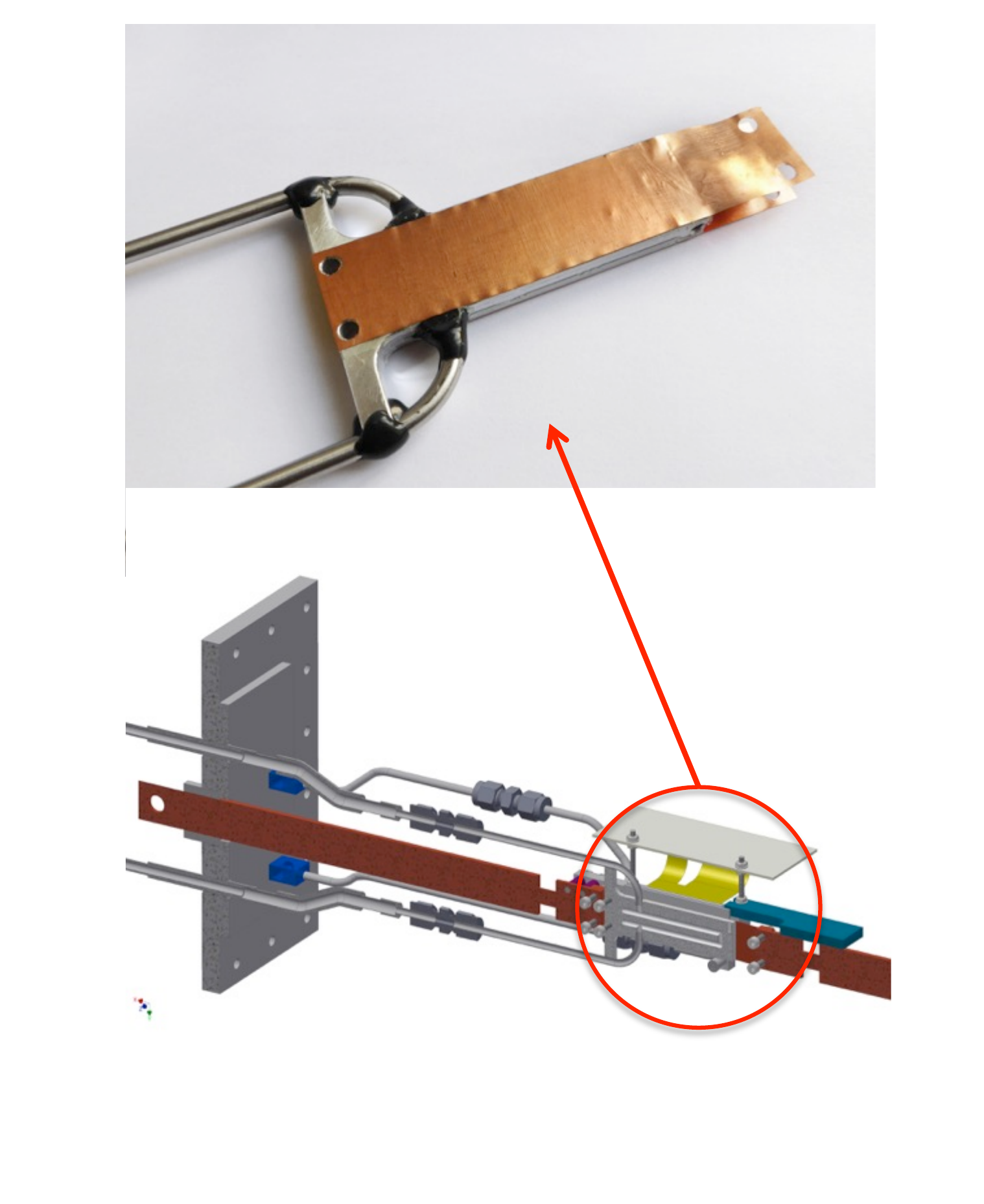}\hspace{2pc}
\caption{Copper target of VIP2. The water-cooled target is designed for currents up to \mbox{200 A.} It is viewed by SDDs from the side.}
\label{fig:target}
\end{figure}

Instead of the copper cylinder used in VIP in the new VIP2 experiment a water-cooled copper foil is used. This target (see fig. \ref{fig:target}) is viewed by the X-ray detectors from both sides.
In VIP2 6 silicon drift detectors (SDDs) \cite{marton} with an active area of 100 mm$^{2}$ each are used as X-ray detectors. The energy resolution at the interesting energy region around 8 keV is comparable or even better compared with that of CCDs. This X-ray detector was very successfully working in the SIDDHARTA experiment at DAFNE which studied the X-ray spectrum of kaonic hydrogen with high precision \cite{siddharta}.
The use of silicon drift detectors (SDDs) provides the opportunity for active shielding of the experiment, because SDDs have a time resolution of about 400 ns. Therefore background events in the detector can be identified with the help of an array of plastic scintillators around the target.

\subsection{Preparation of the VIP2 setup}

Many steps are necessary to set the VIP2 apparatus up and to ensure a stable operation in the underground laboratory, like testing of the target system, the X-ray detectors and the shielding.\\

\subsubsection{Target}
A crucial part of VIP2 is the copper target system. This water-cooled target has to be suitable for long-term running at a current higher than 100 A. The target was already successfully tested up to 180 A. The water cooling system was effective to keep the temperature of the copper foil at room temperature level, which is satisfactory regarding that the end contacts of the copper supply rod are outside the vacuum chamber of the setup. The well-controlled target temperature is crucial since the SDDs are only 10 mm separated from the copper foil and the SDDs have to be cooled to obtail good energy resolution.

\subsubsection{SDDs}
\begin{figure}[h]
\centering
\includegraphics[width=0.9\textwidth]{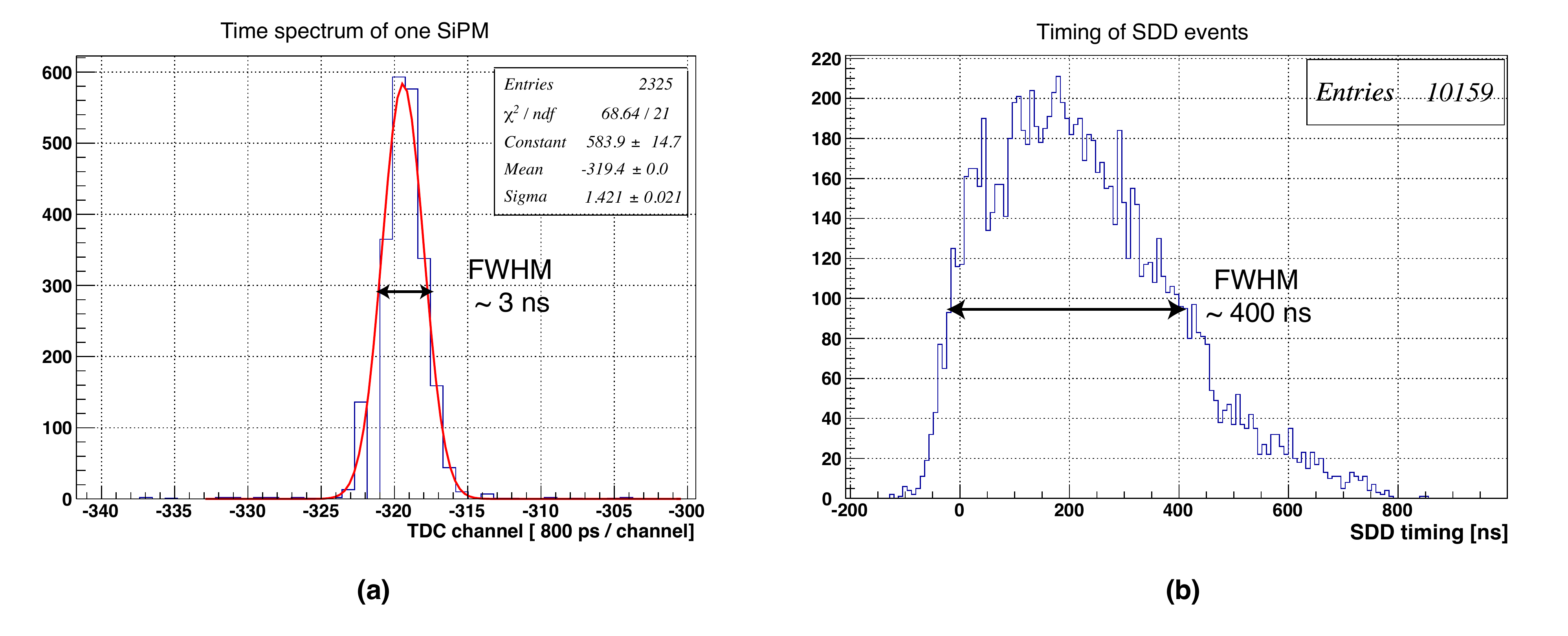}\hspace{2pc}
\caption{Timing performance of one scintillator read out by SiPMs (a) and the time resolution of a typical SDD (b).}
\label{timing}
\end{figure}

An important issue is the energy resolution of the SDD X-ray system. A dedicated Fe-55 source of low radioactivity (about 1 kBq) was used in tests and it was confirmed that this source is suitable for an in-situ calibration source inside the VIP2 setup.
An important feature of SDDs is their timing capability which enables the active shielding with plastic scintillators. The timing performance of SDDs and scintillators with SiPM readout was tested and found to be sufficient for the active shielding (see fig. \ref{timing}).

\subsubsection{Active shielding system}
The active shielding system consists of 32 plastic scintillators readout by 64 silicon photomultipliers (SiPMs). The output signal will be the linear sum of 2 analog signals from the SiPMs. The electronic scheme for the active shielding system was developed and tested.

\begin{figure}[h]
\centering
\begin{minipage}{0.45\textwidth}
\centering
\includegraphics[width=14pc]{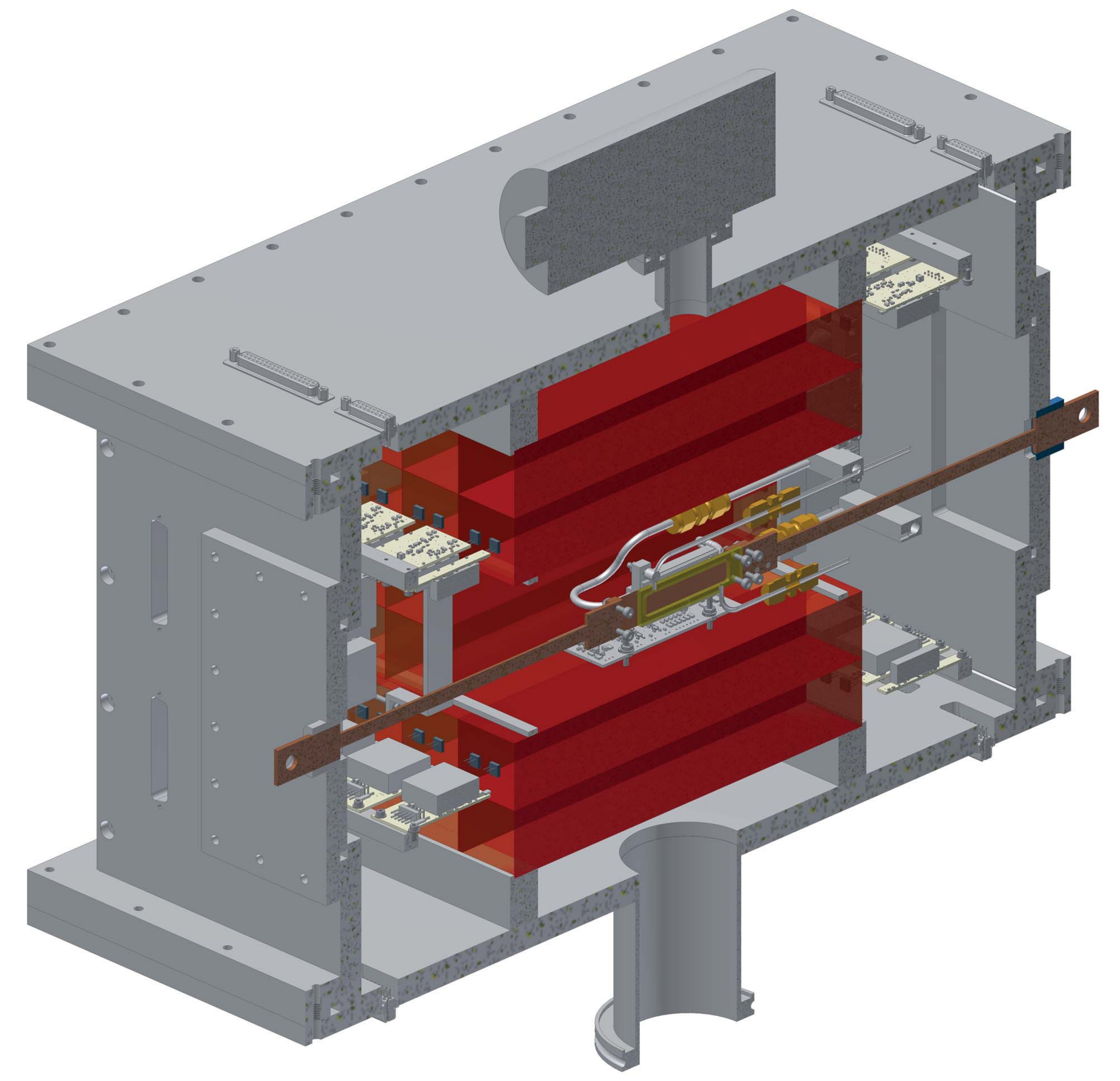}
\caption{Artist's view of the VIP2 setup incorporating the copper target and SDD X-ray detectors. An active shielding system (in red) surrounds the target. The whole inner setup is mounted inside an insulation vacuum box.}
\label{VIP2}
\end{minipage}\hspace{1cm}%
\begin{minipage}{0.45\textwidth}
\centering
\includegraphics[width=10pc]{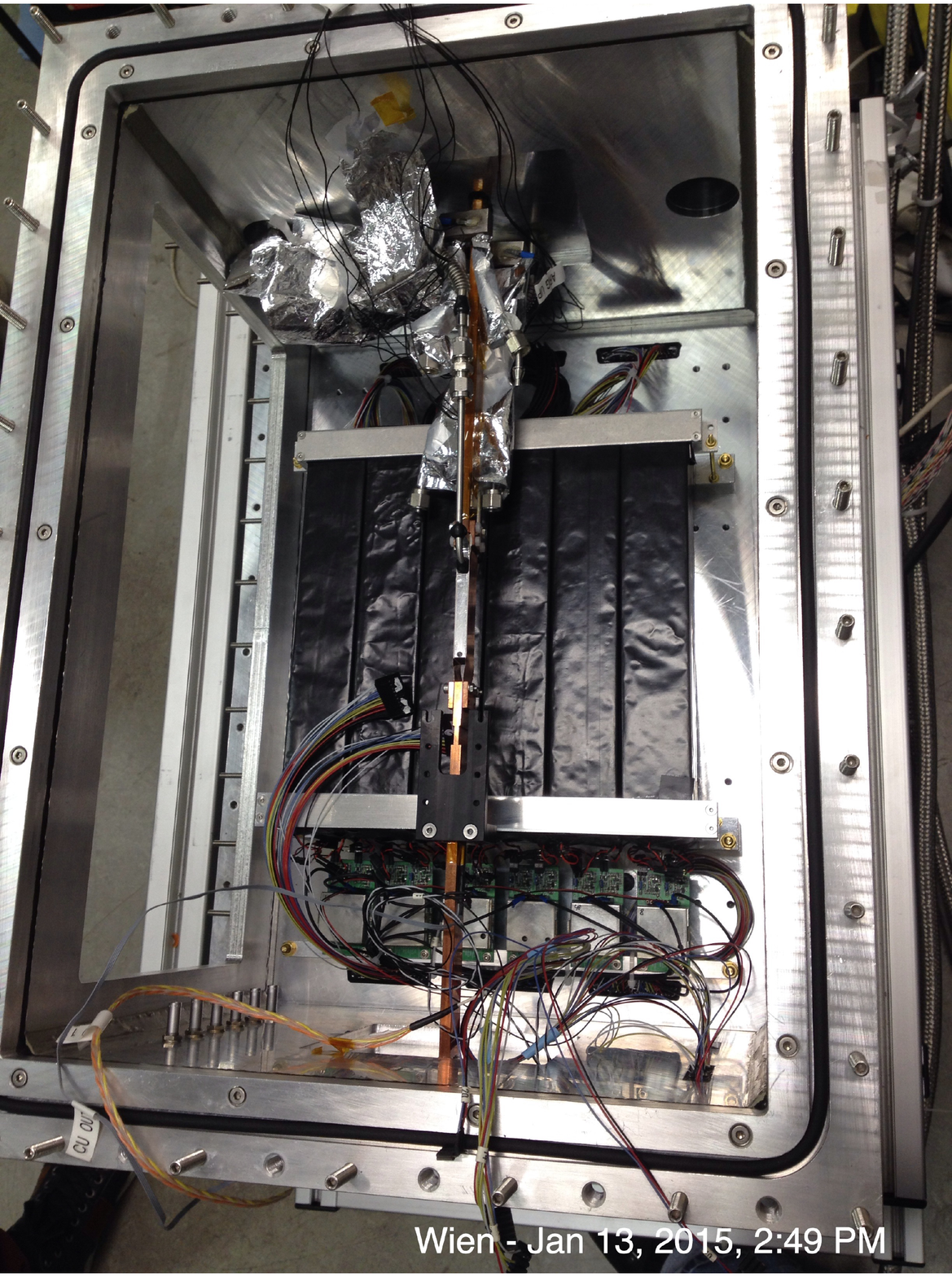}
\caption{Photo of the VIP2 setup. One can see the copper target in the middle and scintillators of the active shielding system at the bottom of the vacuum box.}
\label{box-open}
\end{minipage}
\end{figure}

The designed VIP2 setup is shown in fig. \ref{VIP2}. The vacuum box with the inner part is shown in fig. \ref{box-open}. \\

\section{Outlook}

The history of the Pauli Exclusion Principle and the spin statistics connection is already an old one. It is fascinating that it looks that nature only knows two classes of systems: fermions and bosons. The important decisive criterion is the spin which can only be described in the framework of quantum mechanics. A possible violation of PEP cannot be quantified neither can systems be identified in which one should search for PEP violation. Compared to combined systems like the nucleon, the electron is a very fundamental and clean case, being point-like and furthermore the lightest of the leptons. Therefore high sensitivity searches for tiny PEP violations via X-rays are interesting and experimental methods like the RS method were developed. Due to the advances in instrumentation and detector technology the sensitivity can be increased to a large extent. Therefore low-background high precision experiments have the potential for discoveries - maybe also in the case of PEP.


\subsection{Acknowledgments}

We thank the Austrian Science Foundation (FWF) which supports the VIP2 project with the grant P25529-N20, and we thank the support from the EU COST Action MP1006,
Fundamental Problems in Quantum Physics, and from Centro Fermi (”Problemi aperti nella meccania quantistica”)
project.


\newpage

\section*{References}
\begin{thebibliography}{9}

\bibitem{pauli25} Pauli W. 1925 \emph{Z. Physik} {\bf 31} 765 

\bibitem{pauli46} Pauli W. Nobel prize lecture (1946), http://www.nobelprize.org.

\bibitem{pauli40} Pauli W. 1940 \emph{Phys. Rev.} \textbf{58} 716 

\bibitem{bernabei97} Bernabei R. et al. 1997 {\it Phys. Lett. B}  {\bf 408} 439 

\bibitem{bernabei09} Bernabei R. et al. 2009 New search for processes violating the Pauli exclusion principle in sodium and in
iodine. \textit{Eur. Phys. J.}  \textbf{C 62}, 327 , Bernabei R. et al. 2010 \textit{Journal of Physics: Conference Series} \textbf{202}  012039

\bibitem{borexino05} Borexino~Collaboration, Back H. O. et al. 2005 {\it Eur. Phys. J.} {\bf C37} 421 

\bibitem{borexino10} Bellini G. et al. 2010 New experimental limits on the Pauli forbidden transitions in 12C nuclei obtained
with 485 days Borexino data. \textit{Phys. Rev.} \textbf{C 81}, 034317 

\bibitem{messiah64} Messiah, A.M.L., Greenberg, O.W. 1964 Symmetrization postulate and its experimental foundation. \textit{Phys.
Rev. 136}, \textbf{B248} 


\bibitem{ramberg90} Ramberg E., Snow G. A. 1990 {\it Phys. Lett. B} {\bf 238} 438

\bibitem{sperandio06} Di Matteo S., Sperandio L., {\it  VIP Note, IR-04, 26 April 2006; The energy shift
  has been computed by P. Indelicato, private communication} 

\bibitem{ignatiev87} Ignatiev A. Yu. and Kuzmin V. A. 1987  {\it Yad. Fiz.} {\bf 46} 786; {\it Sov. J. Nucl. Phys.} {\bf 47} 6; see
also the recent review paper Ignatiev A Yu 2006 {\it Rad. Phys. Chem.} {\bf 75}  2090

\bibitem{jackson08} Jackson, Mark G. 2008 {\it Phys. Rev.} {\bf D 78} 126009

\bibitem{balachandran10} Balachandran A.P. et al. 2010 Non-Pauli transitions from spacetime noncommutativity, \textit{Phys. Rev. Lett.} \textbf{105}, 051601 

\bibitem{vip} The VIP proposal, LNF-LNGS Proposal 2004 http://www.lnf.infn.it/esperimenti/vip

\bibitem{egger} Egger J. P., Chatellard D. and Jeannet E. 1993 {\it Particle World} {\bf 3} 139

\bibitem{DEAR} Ishiwatari T. et al. 2004 {\it Phys. Lett. B} {\bf 593} 48; Beer G {\it et al.} 2005 {\it Phys. Rev.
    Lett.} {\bf 94}  212302

\bibitem{ishiwatari} Ishiwatari T. et al. 2006 {\it Nucl. Instrum. Methods Phys. Res. A} {\bf 556}  509 

\bibitem{elliott} Elliott S.R., LaRoque B.H., Gehman, V.M., Kidd M.F. and Chen M. 2012 {\it Foundations of Physics} {\bf 42} 1015-1030 

\bibitem{vipart} Bartalucci S. et al. (VIP~Collaboration) 2006  {\it Phys. Lett. B} {\bf 641} 18

\bibitem{laura} Sperandio L. Ph D thesis New experimental limit on the Pauli Exclusion
Principle violation by electrons from the VIP experiment at University "Tor
Vergata", Roma, 5 March 2008

\bibitem{curceanu11} Curceanu C. et al. 2011 \textit{Journal of Physics} \textbf{306}, 012036 


\bibitem{Hexi14} Hexi S. Testing the Pauli Exclusion Principle for electrons at LNGS, arXiv:1405.1634v1 [quant-ph] 2014, Physics Procedia (20215) in print

\bibitem{Curceanu14} Curceanu C. et al. 2014 Quantum explorations: from the waltz of the Pauli exclusion principle to the rock of the spontaneous collaps \textit{Phys. Scr.} \textbf{90}  028003

\bibitem{Pisicchia15} Pisicchia K. et al. 2014 \textit{Phys. Scr.} \textbf{90} 028003, arXiv[quant-ph]501.04462v1, 2015

\bibitem{Pisicchia15a} Piscicchia K. et al. 2015 Beyond quantum mechanics? Hunting the ‘impossible’ atoms (Pauli Exclusion Principle violation and spontaneous collaps of the wave function at test, arXiv:1501.04462v1 [quant-ph], to be published.

\bibitem{marton} Marton J. et al. 2009 {\it Trans. Nucl. Sci.} {\bf 56} 1400

\bibitem{siddharta} Curceanu C. et al 2007  {\it Eur Phys J} {\bf A31} 537-539;  Bazzi M. et al.  2009 {\it Phys. Lett.} {\bf B 681}  310



  



























































\end{thebibliography}
\end{document}